\documentclass[aip,pop,reprint]{revtex4-1}
\usepackage{tabularx}
\usepackage{amssymb}
\usepackage{amsmath}
\usepackage{amsthm}
\usepackage{nicefrac}
\usepackage{graphicx}
\usepackage{epstopdf}
\usepackage[bookmarks,colorlinks,breaklinks]{hyperref}
\hypersetup{linkcolor=blue,citecolor=blue,filecolor=black,urlcolor=blue}

\makeatletter
\newcommand*{\citenumns}[2][]{%
  \begingroup
  \let\NAT@mbox=\mbox
  \let\@cite\NAT@citenum
  \let\NAT@space\NAT@spacechar
  \let\NAT@super@kern\relax
  \renewcommand\NAT@open{}%
  \renewcommand\NAT@close{}%
  \cite[#1]{#2}%
  \endgroup
}
\makeatother

\begin{document}

\title{Bulk hydrodynamic stability and turbulent saturation in compressing hot spots}

\author{Seth Davidovits}
\affiliation{Princeton University, Princeton, New Jersey 08544, USA}
\author{Nathaniel J. Fisch}
\affiliation{Princeton University, Princeton, New Jersey 08544, USA}
\affiliation{Princeton Plasma Physics Laboratory, Princeton, New Jersey 08544, USA}

\begin{abstract}
For hot spots compressed at constant velocity, we give a hydrodynamic stability criterion that describes the expected energy behavior of non-radial hydrodynamic motion for different classes of trajectories (in $\rho R$ --- $T$ space). For a given compression velocity, this criterion depends on $\rho R$, $T$, and $\mathrm{d}T/\mathrm{d}(\rho R)$ (the trajectory slope), and applies point-wise, so that the expected behavior can be determined instantaneously along the trajectory. Among the classes of trajectories are those where the hydromotion is guaranteed to decrease, and those where the hydromotion is bounded by a saturated value. We calculate this saturated value, and find the compression velocities for which hydromotion may be a substantial fraction of hot-spot energy at burn time. The \citet{lindl1995} ``attractor'' trajectory is shown to experience non-radial hydrodynamic energy that grows towards this saturated state. Comparing the saturation value to available detailed 3D simulation results, we find that the fluctuating velocities in these simulations reach substantial fractions of the saturated value.
\end{abstract}

\maketitle

\section{Introduction}\label{sec:introduction_stability}
Non-radial hydrodynamic motion in the hot spots (gas fill) of inertial-fusion experiments may be seeded by interfacial instabilities (e.g. Rayleigh-Taylor or Richtmeyer-Meshkov instabilities), or by implosion asymmetry generated by a variety of possible sources\citep{weber2015,haines2016}. More broadly, other mechanisms capable of generating non-radial and/or turbulent flow may be at play in compression experiments; experiments in gas-puff Z-pinches suggest significant, and likely turbulent, non-radial hydrodynamic motion at stagnation \citep{kroupp2011,maron2013,kroupp2018}, which may be generated and carried along during the compression itself. Here, non-radial hydrodynamic motion refers to motion not associated with the compression itself; this motion may be regarded as ``wasted energy'' to the extent it does not contribute to heating in the stagnation process, and may also degrade performance\citep{hammel2011,regan2012,ma2013,haines2016}, for example, through inducing mix of colder or non-fuel capsule material into the hot-spot.
On the other hand, it may be possible to design a new type of fast-ignition scheme that uses large quantities of such hydrodynamic motion to spark fusion or a burst of X-rays \citep{davidovits2016a,davidovits2016b}. It is also the case that large hydrodynamic motion affects the interpretation of spectroscopic measurements, beyond Doppler-broadening effects; density fluctuations may be induced, which must be accounted for self-consistently for a correct treatment \citep{kroupp2018}.

In either case, whether one is interested in reducing or utilizing such hydrodynamic motion, it is difficult to predict how much hydrodynamic motion will be present in a given experiment, or even, grossly, whether one experiment would be expected to have more or less of such motion than another experiment. Currently, determining the expected amount of non-radial hydrodynamic motion in an implosion requires very computationally expensive and time consuming three-dimensional (3D) simulations (e.g. Refs. \citenumns{thomas2012,weber2014,weber2015,clark2015,haines2016}). Although such simulations, as the most inclusive accounting of the implosion dynamics, will always have their place, it is desirable, due to their expense, to also develop less expensive methods that can serve as a first level of examination. This first level of examination can then be used to help determine when detailed 3D modeling is more likely needed from a perspective of hot-spot (gas-fill) hydrodynamics. It is also desirable to develop better intuition about the behavior of non-radial hydrodynamic motion in such experiments from a design perspective.

As a step towards these ends, a model (ordinary differential equation) that predicts the (turbulent) non-radial hydrodynamic motion (hydromotion) for plasma undergoing 3D, constant velocity, compression has recently been developed \citep{davidovits2017b}. Other modeling efforts provide more detail on the influence of initial conditions \cite{viciconte2018}. As a separate, parallel approach, the present work develops a stability criterion for hot-spot hydromotion that gives intuition and predictions as to which trajectories (in $\rho R$ vs $T$ space) are more or less likely to have substantial hydromotion. Further, we give a saturation level for the hydromotion on certain trajectories, and explain how, for many trajectories, this can be regarded as a limit on how large the hydromotion could possibly get. The stability and saturation results are found to compare favorably with limited accessible results in a comparison to detailed 3D simulations\citep{weber2015,clark2015}, carried out in Sec.~\ref{sec:discussion_stability}. All of this is done in the context of the treatment outlined in Sec.~\ref{sec:treatment_stability}; because the treatment does not necessarily include all possible important effects, the present results may be most useful from an intuitive and gross-estimation sense, rather than as an exact demarcation of stability boundaries or saturation levels. 

Throughout the work, we will use TKE (turbulent kinetic energy) as a shorthand to refer to non-radial hydrodynamic motion (hydrodynamic motion in the frame moving with the compression). For the sake of the stability criterion, whether or not the flows are truly ``turbulent'' (have a well-developed inertial range) is not particularly important (the analysis does however assume isotropy and homogeneity). Similarly, saturation can occur for quite modest Reynolds numbers (simulations in Ref.~\citenumns{davidovits2017b} can saturate at large scale Reynolds numbers on order of 100). 
The word ``bulk'' in the title is meant to emphasize that the effects examined here are due to the volumetric compression; in particular no interfacial instability is necessary, although such instabilities are one possible seed for non-radial hydrodynamic motion. The terms hot spot and gas fill are used interchangeably in this work; in particular, when we refer to the hot spot, there is no temperature requirement implied. Rather, we mean the interior plasma being compressed (by the capsule) throughout the implosion.

The compression velocity in the present work is assumed to be instantaneously constant. By this, we mean that effects from acceleration on the bulk hydromotion are not included; no acceleration is needed for the growth of the hydromotion in the compressions considered here. The results can still be applied to a compression where the implosion velocity changes over the course of the implosion, by recalculating for the new velocity as necessary (keeping in mind acceleration impacts are still neglected).

The work is organized as follows. The following section, Sec.~\ref{sec:treatment_stability}, describes the model underlying the present work. Next, Sec.~\ref{sec:general_stability} gives the stability and saturation results for a general viscosity. Section~\ref{sec:braginskii_stability} specializes the results to the (unmagnetized, parallel) Braginskii viscosity; picking a specific form for the viscosity then allows the results to be displayed visually in $T$ vs $\rho R$ space. The stability and saturation results are displayed in concert with a simple hot-spot model in Sec.~\ref{sec:context}, which provides further context for them. In Sec.~\ref{sec:discussion_stability}, we compare the stability and saturation results to the predictions of detailed simulations of a National Ignition Campaign experiment \citep{weber2015,clark2015}, and discuss caveats and restrictions for the present treatment. Finally, Sec.~\ref{sec:summary_stability} summarizes the main results and concludes.

\section{Treatment} \label{sec:treatment_stability}
To examine TKE behavior in the hot spot, we consider the isotropic, 3D, compression of a plasma modeled as a fluid. The fluid equations are taken to be the Navier-Stokes (NS) equations, with a viscosity that depends on time (equivalently, compression ratio). The plasma flow in the hot-spot is broken into two components, a mean compressive flow towards the origin, $\mathbf{v}_0$, and the fluctuating flow, $\mathbf{v}'$. The compressing flow is taken as a given (enforced), while we will solve for the evolution of the fluctuating flow from some initial state. The compressing flow is
\begin{equation}
\mathbf{v}_0 = \frac{\dot{L}}{L} \mathbf{x}, \label{eq:compressing_flow}
\end{equation}
where $L$ is defined,
\begin{equation}
L (t) = L_0 - 2 U_b t . \label{eq:L}
\end{equation}
The overdot in Eq.~(\ref{eq:compressing_flow}) indicates the time derivative; note that $\dot{L}$ is negative, so that the flow $\mathbf{v}_0$ is compressing (has negative divergence). In Eq.~(\ref{eq:L}), $L_0$ is the initial side length of the domain, and $U_b$ is a compression velocity. The effect of the flow $\mathbf{v}_0$ can be described as follows. A cube of side length $L_0$, placed in the flow $\mathbf{v}_0$ at $t=0$, and advected by flow, will remain a box and shrink in time, with the side length given by $L (t)$ in Eq.~(\ref{eq:L}). The (constant) velocity of the box sides is then $U_b$. The (linear) compression ratio, $\bar{L}$, is given by $\bar{L} = L/L_0$. 

We will further assume that the fluctuating flow, $\mathbf{v}'$, is low (zero) Mach, so that we can ignore sound waves and any density perturbations. In this case, $\mathbf{v}'$ is incompressible (divergence free). Since we are ignoring density fluctuations, the continuity equation gives that the density, $\rho$, increases as expected for a 3D, isotropic compression,
\begin{equation}
\rho (t) = \frac{\rho_0}{\bar{L}^3}. \label{eq:density}
\end{equation}
Similarly, we will assume that the hot-spot temperature is spatially uniform, allowing the temperature to only depend on the compression ratio (time), $T = T (\bar{L})$. 

With these assumptions, and working in a frame moving with the mean compressive flow, the NS momentum equation can be written,
\begin{equation}
\frac{\partial \mathbf{V}}{\partial t} + \frac{1}{\bar{L}} \mathbf{V} \cdot \nabla \mathbf{V} - \frac{2 U_b}{L} \mathbf{V} + \frac{\bar{L}^2}{\rho_0} \nabla P = \frac{\mu_0}{\rho_0} \bar{\mu} (\bar{L}) \bar{L} \nabla^2 \mathbf{V}. \label{eq:momentum}
\end{equation}
Here $\mathbf{V}$ is the fluctuating velocity rewritten in the moving coordinates, $\mathbf{V} (\mathbf{X},t) = \mathbf{v} (\mathbf{x},t)$, where the transformation to the moving frame is $\mathbf{x} = \bar{L} \mathbf{X}$. The dynamic viscosity, $\mu$, is written as $\mu_0 \bar{\mu} (\bar{L})$, where $\bar{\mu} (\bar{L} = 1) = 1$, so that the initial viscosity is $\mu_0$. The viscosity is described further below.
A more complete derivation of the preceding is given in the Appendix of \citet{davidovits2016b}, although we use slightly different assumptions surrounding the temperature behavior here (which enters through the viscosity); very similar models for compressing gas/fluid have been used elsewhere as well \citep{wu1985,coleman1991,blaisdell1991,cambon1992,hamlington2014,robertson2012,peebles1980}.

In general, the viscosity $\mu$ may depend on properties of the plasma, for example the temperature or the charge state, $Z$. As such, the viscosity can vary during the compression. Without losing generality, we can say that, for a given experiment or simulation, the viscosity is some (fit) function of compression, $\mu (\bar{L}) = \mu_0 \bar{\mu} (\bar{L})$. 

Later, we will present some results where we have specialized to the (parallel, unmagnetized) Braginskii viscosity. The Braginskii viscosity depends on the plasma charge state and temperature as $\mu \sim T^{5/2}/Z^4$. As with the overall viscosity, generally speaking, for any given compression, the charge state of the hot spot could be written as some function of compression $Z = Z (\bar{L})$; this could be achieved through a model, or regarded as a fit to experimental or simulation results. The same can be said for the temperature. Then, the viscosity can be regarded as a function of $\bar{L}$. 

\section{General stability and saturation} \label{sec:general_stability}
The basic result underlying the stability criterion is derived in \citet{davidovits2016b}, Sec.~III. Here, we discuss the application of this result to hot-spot TKE, and recast the result in a form that is more useful for this purpose. 

The energy density of the fluctuating flow is $E = \rho_0 \mathbf{V}^2/2$. The total energy, $E^T$, is then the integral of the energy density over the domain, or, equivalently, integrated over all Fourier modes, 
\begin{equation}
E^T = \int_{k_{\rm{min}}}^{\infty} dk E (k,t),
\end{equation}
where the minimum wavenumber is set by $L_0$, $k_{\rm{min}} = 2\pi/L_0$. An equation for the time evolution of this total energy can be written making use of the momentum equation, Eq.~(\ref{eq:momentum}). By demanding that all Fourier modes of the TKE are damped (linearly), \citet{davidovits2016b} arrive at a condition that is sufficient to guarantee the total hot-spot TKE will decrease as the hot-spot is compressed. Written using the dynamic viscosity, this TKE decrease condition is
\begin{equation}
\frac{ U_b \rho L}{2 \mu} < \pi^2. \label{eq:decrease_cond}
\end{equation}
Note that this condition is derived for a cubic domain, while hot-spots are typically spherical; this will introduce some small error, but the analysis, in using a limited model, is intended only as a general guide anyway.

The left hand side of Eq.~(\ref{eq:decrease_cond}) is essentially a Reynolds number, with the velocity the compression velocity, and the length scale half the domain length (the ``radius'', ignoring geometric factors). In general, this Reynolds number will change as the compression progresses; the areal density $\rho L/2$ will increase (as $\bar{L}^{-2}$), while the viscosity may increase or decrease. The left hand side of Eq.~(\ref{eq:decrease_cond}) will be a constant for the special case where $\bar{\mu} = \bar{L}^{-2}$, in which case the viscosity compression dependence cancels the dependence in the areal density. In this case, the inequality will be either satisfied or not satisfied for the entire compression.

If the inequality, Eq.~(\ref{eq:decrease_cond}), is satisfied for the full duration of the compression, $\bar{L} \in [1,\bar{L}_{\rm{final}}]$, then we can say that the final TKE will be lower than its initial value, as the TKE decreases throughout the compression. If the viscosity behavior for a compression is known, from either a model, simulation, or experiments, this gives a simple (energy) ``stability'' criterion. Note that the condition applies point-wise along the compression; if it is satisfied at some values of $\bar{L}(t)$, but not others, this determines when during the compression the TKE is guaranteed to decrease. When the condition, Eq~(\ref{eq:decrease_cond}), is not satisfied, the TKE may increase, decrease, or stay the same, depending on its present value. That is, when the inequality is not satisfied, the TKE behavior is not uniquely determined by the inputs of Eq.~(\ref{eq:decrease_cond}).

If $\mu$ depends only on the hot-spot temperature (and/or the areal density), then the TKE decrease condition, Eq.~(\ref{eq:decrease_cond}), depends only on $\rho L$ and $T$ for a given $U_b$. In this case, the TKE decrease condition can be plotted as a ``stability boundary'' in $\rho R - T$ space. This will be done for the Braginskii viscosity in Sec.~(\ref{sec:braginskii_stability}). 

In the event that $\bar{\mu} (\bar{L}) = \bar{L}^{-2}$, and Eq.~(\ref{eq:decrease_cond}) is not satisfied, it can be shown that the TKE will change under compression towards a saturated value, which it will reach for a sufficient amount of compression. The turbulent energy density in this saturated state is\citep{davidovits2016b},
\begin{equation}
E_{\rm{sat}} = 1.9 \rho U_b^2. \label{eq:saturated_energy_density}
\end{equation}
Note that, the total hot-spot mass is conserved in the present model, which means that the saturated total TKE is in fact a constant for fixed $U_b$; the total TKE, $E_T$, is $E_T \propto r^3 E_{\rm{sat}} \propto r^3 \rho U_b^2 \propto m_{\rm{hotspot}} U_b^2$, with $m_{\rm{hotspot}}$ the hot-spot mass. As an example, when $U_b = 3\times10^7$ cm/s, a hot-spot with a mass of 800 ng has a saturated TKE of $\sim140$ J.

Consider a hot-spot undergoing compression for which $\bar{\mu} (\bar{L}) = \bar{L}^{-2}$ and the TKE decrease condition, Eq.~(\ref{eq:decrease_cond}) is not satisfied. In this case, if the TKE density is larger than $E_{\rm{sat}}$, the TKE will decrease with compression. On the other hand, if the TKE density is below this value, it will increase towards the saturated value as the hot-spot is compressed. A source for TKE outside the present model is needed for the TKE density to exceed $E_{\rm{sat}}$; in other words, if the turbulent velocity is governed by Eq.~(\ref{eq:momentum}), the TKE density for compressions with $\bar{\mu} (\bar{L}) = \bar{L}^{-2}$ will not exceed $E_{\rm{sat}}$.

For compressing hot-spots where the viscosity growth with compression is stronger than $\bar{\mu} (\bar{L}) = \bar{L}^{-2}$ (say, grows as $\bar{L}^{-n}$ with $n>2$), we suggest (but have not proven) that the TKE energy density will be bounded above by Eq.~(\ref{eq:saturated_energy_density}). This follows from the arguments underlying the bound in \citet{davidovits2017a}; essentially, a stronger viscosity growth should not lead to less dissipation than in an identical compression with a weaker viscosity growth. It may also be the case that, even for weaker viscosity growth (say, $n < 2$ in the previous expression for $\bar{\mu}$), the TKE density will not exceed $E_{\rm{sat}}$; this is the result given by a TKE model \citep{davidovits2017b}.

\begin{figure}
\includegraphics[width = \columnwidth]{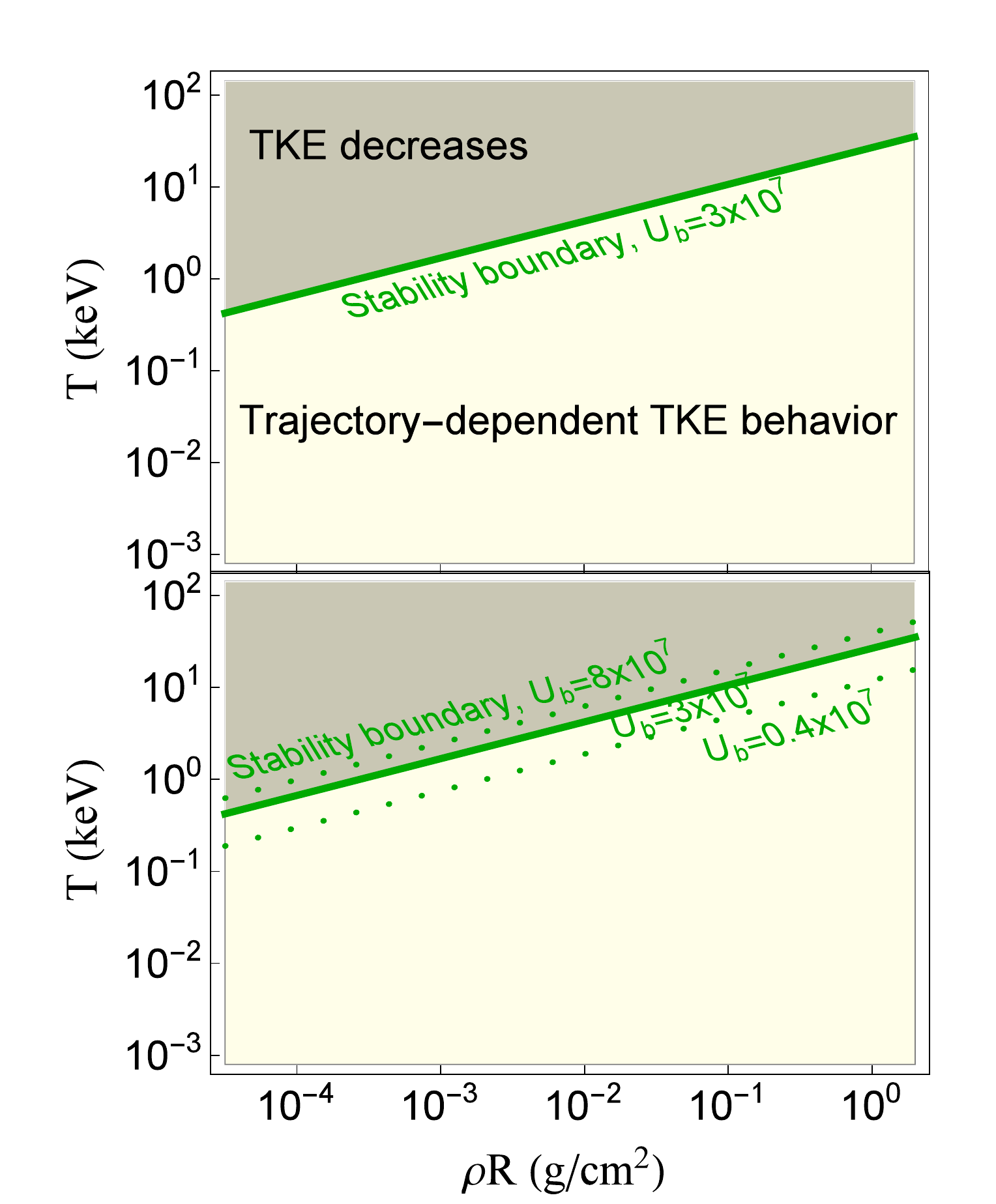}
\caption{TOP: Visual representation of the stability boundary, Eq.~(\ref{eq:stability_boundary}), in $T$ vs $\rho R$ space. This is the specialization of the general TKE decrease condition, Eq.~(\ref{eq:decrease_cond}), to the case with Braginskii viscosity. The boundary is the labeled, solid, green line; it is shown for an implosion velocity of $U_b = 3 \times 10^{7}$ cm/s, and assuming $A_i = 2.5$ and $\ln \Lambda = 2$. The darker shaded region, above the line, is the region where hot-spot TKE will decrease even with forcing from the compression. The hot-spot TKE behavior in the lighter shaded region, below the line, depends on, among other things, the hot-spot trajectory slope. See Secs.~\ref{sec:general_stability} and \ref{sec:braginskii_stability}, as well as Table~\ref{tbl:cases}.
BOTTOM: Stability boundary for two additional implosion velocities, $U_b=8\times10^7$ cm/s and $U_b=0.4\times10^7$ cm/s, as well as that for $U_b =3\times10^7$ cm/s. The region shading is still for the $U_b=3\times10^7$ cm/s implosion velocity boundary. As the implosion velocity increases, the region of guaranteed TKE decrease shrinks, requiring higher hot-spot temperatures at a given $\rho R$.
\label{fig:stability_plot}}
\end{figure}

\begin{figure}
\includegraphics[width = \columnwidth]{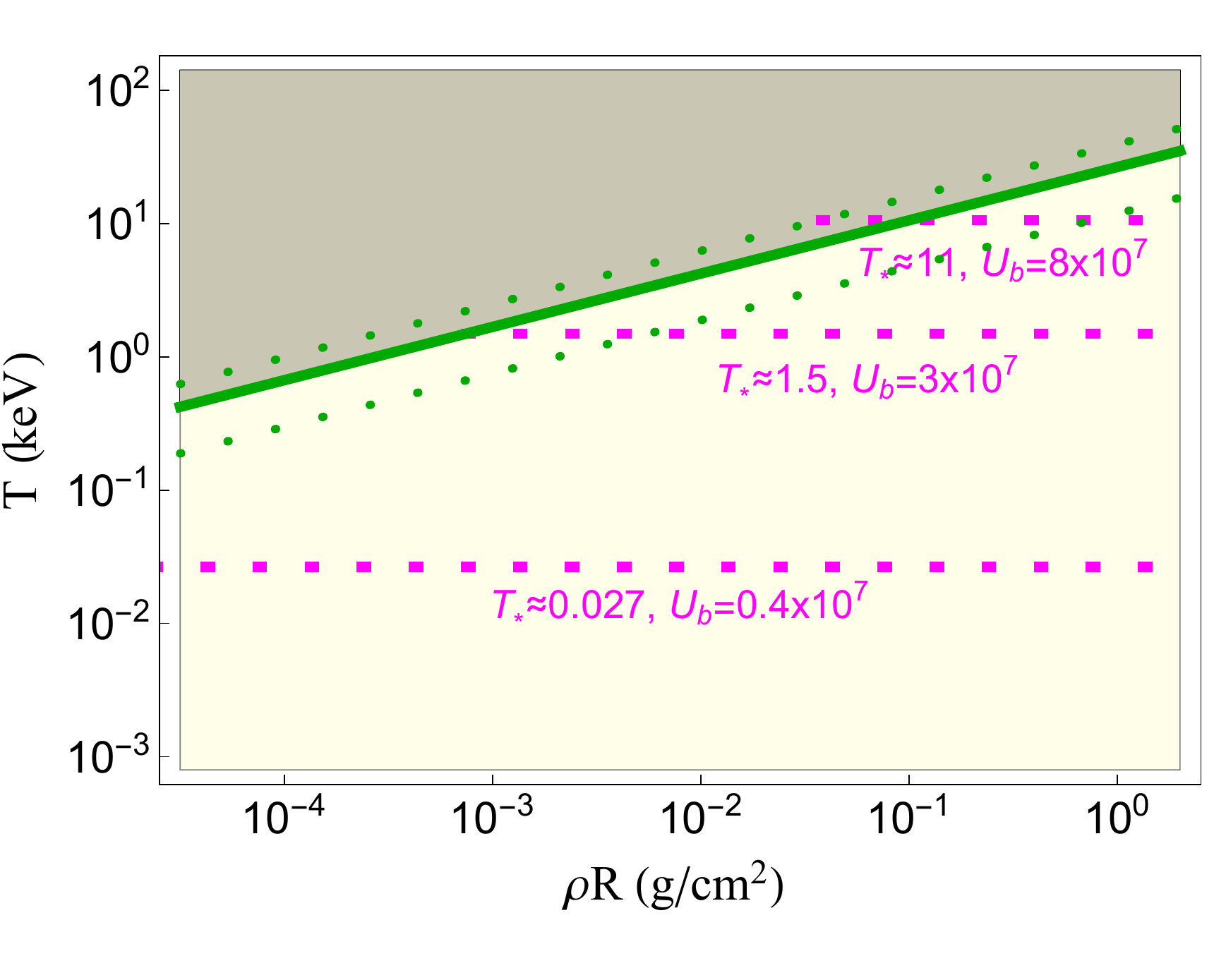}
\caption{Values of $T_{*}$ for three different compression velocities, plotted in $T$ vs $\rho R$ space, where they are horizontal lines (dotted, magenta). Also plotted are the stability boundaries for these three velocities, and stability shading, as in Fig.~\ref{fig:stability_plot}. In the region below the stability boundary, the hot-spot TKE density in the present model is generally restricted to be less than or equal to $E_{\rm{sat}}$, Eq.~(\ref{eq:saturated_energy_density}). By calculating the ratio of $E_{\rm{sat}}$ to the hot-spot thermal energy, we can determine where in $T$ vs $\rho R$ space the hot-spot thermal energy will necessarily exceed any TKE. The temperature for which this ratio, Eq.~(\ref{eq:energy_ratio}), is $1$, is defined to be $T_{*}$. For compression velocities ranging from $U_b = 4 \times 10^6$ cm/s to $U_b = 8 \times 10^7$ cm/s, the we find $T_{*}$ to range from 270 eV at the slowest velocity, to 11 keV at the fastest. This means the TKE can either be necessarily small or possibly substantial at fusion temperatures of $\sim10$ keV, depending on the compression velocity. This analysis assumes the initial TKE is below $E_{\rm{sat}}$; see Sec.~\ref{sec:general_stability} for more discussion.
\label{fig:saturation_plot}}
\end{figure}

Since $E_{\rm{sat}}$ acts as either a saturated or bounding TKE density during hot-spot compressions, depending on the rate of viscosity change with compression, it is instructive to compare it to the hot-spot thermal energy density. This tells us, in a sense, how ``bad'', or, in other words, how large, the TKE could possibly get, as a fraction of hot-spot energy, assuming we start with the TKE small and the TKE is forced only by the volumetric compression itself (as it is here). Using Eq.~(\ref{eq:saturated_energy_density}), it is easy to calculate the ratio of thermal energy density, $E_{\rm{th}}$, to TKE density. Assuming a 50/50 deuterium-tritium plasma, and counting the electron thermal energy, the thermal energy density is $E_{\rm{th}} = 3 n k_b T$, with n the (combined) ion number density. Then, using $\rho = 2.5 m_p n$, with $m_p$ the proton mass, the ratio can be written,
\begin{equation}
\frac{E_{\rm{th}}}{E_{\rm{sat}}} = 0.67 \left(\frac{3\times10^7}{U_b} \right)^2 T_{\rm{keV}}. \label{eq:energy_ratio}
\end{equation}
Here $U_b$ is in cgs units, and $T$ is in kilo-electron volts. Given a compression velocity, there is some temperature, $T_{*}$, for which $E_{\rm{th}} = E_{\rm{sat}}$. This temperature is plotted (as a horizontal line in $T$ vs $\rho R$ space) in Fig.~\ref{fig:saturation_plot} for a few different compression velocities. For temperatures an order of magnitude above this value ($T \sim 10 T_{*}$), the thermal energy will dominate even saturated TKE, thus guaranteeing that, within the current treatment, the TKE is a small fraction of hot-spot energy. For a typical compression velocity, $U_b = 3\times10^7$ cm/s, we find $T_{*} \approx 1.5$ keV. Then, for a capsule reaching an ignition temperature of $\sim10$ keV, the thermal energy will necessarily exceed the TKE (again, if the TKE starts below $E_{\rm{sat}}$), but the TKE can, in theory, still be a substantial fraction ($\sim13$\%) of hot-spot energy. The energy ratio, Eq.~(\ref{eq:energy_ratio}), is quite sensitive to compression velocity, and for lower velocity implosions, $T_{*}$ quickly moves into the 10s or 100s of eV. Note that, even if the TKE is small compared to thermal energy, it could still have important effects.

In the following section, we specialize the results in this section to the Braginskii viscosity, showing how the stability results can be applied visually in $T$ vs $\rho R$ space. Then, in Sec.~(\ref{sec:context}), we present the stability and saturation results in the context of a simple hot-spot model, such as one in \citet{lindl1995} or \citet{atzeni2004}. This allows for more concrete discussion. As part of that presentation, we show that the case $\bar{\mu} (\bar{L}) = \bar{L}^{-2}$ is not just a curiosity; in the simple model, it is the viscosity dependence for a hot-spot with mechanical heating balancing electron thermal conduction. The stability and saturation results could similarly be specialized for other viscosity models.

\section{Braginskii viscosity stability and saturation} \label{sec:braginskii_stability}
\begin{table}
\caption{Breakdown of stability and saturation cases, based on trajectory $T$, $\rho R$, and slope. Results given are for the case of Braginskii viscosity with no ionization, but an identical set of cases exists in general, with the conditions placed instead directly on the viscosity behavior. See the discussions in Secs.~\ref{sec:general_stability} and \ref{sec:braginskii_stability}. These cases allow a visual identification of hot-spot TKE behavior once the hot-spot trajectory is plotted in $T$ vs $\rho R$ space. For cases B1 and B2, the degree to which $E_{\rm{sat}}$ is reached during the compression will depend on not only the trajectory in $T$ vs $\rho R$ space, but also the values of $U_b$, $E_0$ (the initial TKE), and (to some degree) the initial Fourier spectrum of the TKE. Similarly for case B3; here, the TKE will typically have a maximum that is only a fraction of $E_{\rm{sat}}$ \citep{davidovits2016b}. Given identical starting conditions, shallower slopes (weaker viscosity growth with compression) correspond with quicker TKE growth per compression increment.} \label{tbl:cases} 
\begin{tabularx}{\linewidth}{ l X }
\hline
A.   & $T(\rho R) > T_{\rm{boundary}} (\rho R)$ \\ \\
\multicolumn{2}{p{\linewidth}}{Trajectory at $\rho R$ is above the stability boundary. Total hot-spot TKE is instantaneously decreasing \citep{davidovits2016b}.} \\
\hline
B.   &  $T(\rho R) < T_{\rm{boundary}} (\rho R)$ \\ \\
\multicolumn{2}{p{\linewidth}}{Trajectory at $\rho R$ is below the stability boundary. TKE behavior depends on trajectory} \\
\hline
B1. & B \& $\frac{\rm{d}T}{\rm{d}(\rho R)}(\rho R) < \frac{\rm{d}T_{\rm{boundary}}}{\rm{d}(\rho R)_{\rm{boundary}}}(\rho R)$ \\ \\
& Slope shallower than stability boundary. TKE hypothesized to asymptote to $E_{\rm{sat}}$ with sufficient compression. 
\\
\hline
B2. & B \& $\frac{\rm{d}T}{\rm{d}(\rho R)}(\rho R) = \frac{\rm{d}T_{\rm{boundary}}}{\rm{d}(\rho R)_{\rm{boundary}}}(\rho R)$ \\ \\
& Slope parallels stability boundary. TKE aymptotes to $E_{\rm{sat}}$ with sufficient compression \citep{davidovits2016b}. \\
\hline
B3. & B \& $\frac{\rm{d}T}{\rm{d}(\rho R)}(\rho R) > \frac{\rm{d}T_{\rm{boundary}}}{\rm{d}(\rho R)_{\rm{boundary}}}(\rho R)$ \\ \\
& Slope steeper than stability boundary. TKE bounded above by $E_{\rm{sat}}$, assuming initial TKE below $E_{\rm{sat}}$.
\\
\hline
\end{tabularx}
\end{table}

While the TKE decrease condition, Eq.~(\ref{eq:decrease_cond}), is useful, we can get much more insight into hot-spot TKE after specifying the hot-spot viscosity model. This is because specifying the viscosity model allows us to determine how $\mu(\bar{L})$ relates to the temperature behavior $T (\bar{L})$ (and therefore to the trajectory in $\rho R$ vs $T$ space). In the present work, we use the unmagnetized (parallel) Braginskii viscosity, 
\begin{equation}
\mu_{\rm{Brag}} (\bar{L}) = \mu_{0,\rm{Brag}} \frac{\bar{T}^{5/2}}{\bar{Z}^4}. \label{eq:brag_viscosity}
\end{equation}
Here, as elsewhere, the overbar on $T$ and $Z$ indicates normalization to initial values at $L = L_0$ ($\bar{L}=1$), $\bar{T} = T/T_0$, $\bar{Z} = Z/Z_0$. As previously noted, in general $T$ and $Z$ will be some functions of compression. For the work here, we will assume that $Z = \rm{constant}$, so that $\bar{Z} =1$. That is, we assume that there is no change in the ionization state of the hot spot as it compresses. To the extent that there is not substantial ongoing mixing of different Z material (shell) into the hot spot, this is a reasonable assumption, and it should make the discussion easier to follow. The assumption is not fundamental to the present analysis, and can be relaxed, given an expression for $Z (\bar{L})$ (or, say $Z (T)$, with $T(L)$ then modeled, simulated, or measured).

For reasons that will soon become apparent, it is useful to discuss ``hot-spot trajectories''; a trajectory tracks the hot-spot temperature as a function of $\rho R$ (equivalently, $\bar{L}$ or time), starting from some initial point, $T_0, (\rho R)_0$. As such, trajectories are curves in $T$ vs. $\rho R$ space. The slope ($\sim \rm{d}T/\rm{d} (\rho R)$) of the trajectory gives the heating (positive slope) or cooling (negative slope) of the hot-spot with compression. In general, this slope depends on the net balance of a variety of physical processes, for example mechanical ($P \rm{d}V$) work, conduction losses, and radiation losses. As an example, consider a hot-spot during a time when mechanical heating dominates any losses; in this case $\bar{T} = \bar{L}^{-2} = \bar{\rho R}$ (adiabatic heating), and the trajectory would have a slope of $1$ during this time.

Using the Braginskii viscosity, we can reformulate the TKE decrease condition, Eq.~(\ref{eq:decrease_cond}), as a curve in $T$ vs. $\rho R$ space, which then serves as a type of ``stability boundary''. To reformulate the TKE decrease condition, Eq.~(\ref{eq:decrease_cond}), as a stability boundary, we simply substitute in $\mu_{\rm{Brag}}$ for $\mu$, which results in a left hand side that depends only on the (instantaneous) areal density and temperature. Then, we assume equality in the condition, giving,
\begin{equation}
T_{\rm{boundary}} \approx 24.6 \left(\frac{\rm{ln} \Lambda}{A_i^{1/2}} \right)^{2/5} \left( (\rho R)_{\rm{boundary}} \frac{U_b}{3\times10^7} \right)^{2/5}. \label{eq:stability_boundary}
\end{equation}
The substitution $L = 2 R$ has been made, $\rm{ln} \Lambda$ is the Coulomb logarithm, and $A_i$ is the ion atomic mass number. The implosion velocity and $\rho R$ are in cgs units, while the temperature is in kilo-electron volts.  In the present work, we will treat the Coulomb logarithm as a constant. Given values for $A_i$, $\rm{ln} \Lambda$, and $U_b$, equation (\ref{eq:stability_boundary}) is a curve $T_{\rm{boundary}} ( (\rho R)_{\rm{boundary}})$ in $T$ vs. $\rho R$ space. This curve, which is plotted in Fig.~\ref{fig:stability_plot}, represents the marginal case of the TKE decrease condition, and so is a type of stability boundary, the use of which we now describe. 

A trajectory is above the stability boundary if it has $T > T_{\rm{boundary}}$ for $\rho R = (\rho R)_{\rm{boundary}}$. While a trajectory is above the stability boundary, it satisfies the TKE decrease condition, Eq.~(\ref{eq:decrease_cond}), and therefore has decreasing TKE. As such, trajectories that are entirely above the stability boundary will have a final TKE below the initial TKE. Trajectories that cross the stability boundary from above or below will have decreasing TKE for the time they are above the stability boundary; they may or may not have decreasing TKE when below the stability boundary. That is, trajectories below the stability boundary are not necessarily ``unstable'' (do not necessarily experience growing TKE).

As in the general case discussed in Sec.~\ref{sec:general_stability}, there are a few things we can say about the TKE behavior of hot-spots when their trajectories are below the stability boundary, depending on $\bar{\mu}$. As previously discussed, compressions where $\bar{\mu} (\bar{L}) = \bar{L}^{-2}$ have TKE that changes towards a saturated value, $E_{\rm{sat}}$, given in Eq.~(\ref{eq:saturated_energy_density}). Having specialized to the Braginskii viscosity, and still assuming $\bar{Z}=1$, the condition $\bar{\mu} (\bar{L}) = \bar{L}^{-2}$ is equivalent to $\bar{T} = \bar{L}^{-4/5} = \bar{\rho R}^{2/5}$. Of note is that this rate of temperature growth is the same as that on the stability boundary, Eq.~(\ref{eq:stability_boundary}). Thus, trajectories on which the TKE heads towards saturation have the same slope (in $\rm{log}(T)$ vs $\rm{log}(\rho R)$ space) as the stability boundary. The conditions for stronger or weaker viscosity growth, discussed in Sec.~\ref{sec:general_stability}, now correspond to the slope of the trajectory; a slope steeper than the stability boundary slope indicates the TKE can not reach even the saturated value, while a slope shallower than the stability boundary means the TKE may reach at least the saturated value, given enough compression.

The different possible behaviors for the hot-spot TKE, depending on the trajectory slope and location, are summarized in Table \ref{tbl:cases}. 

\section{Stability and saturation: hot-spot model context}\label{sec:context}
\begin{figure}
\includegraphics[width = \columnwidth]{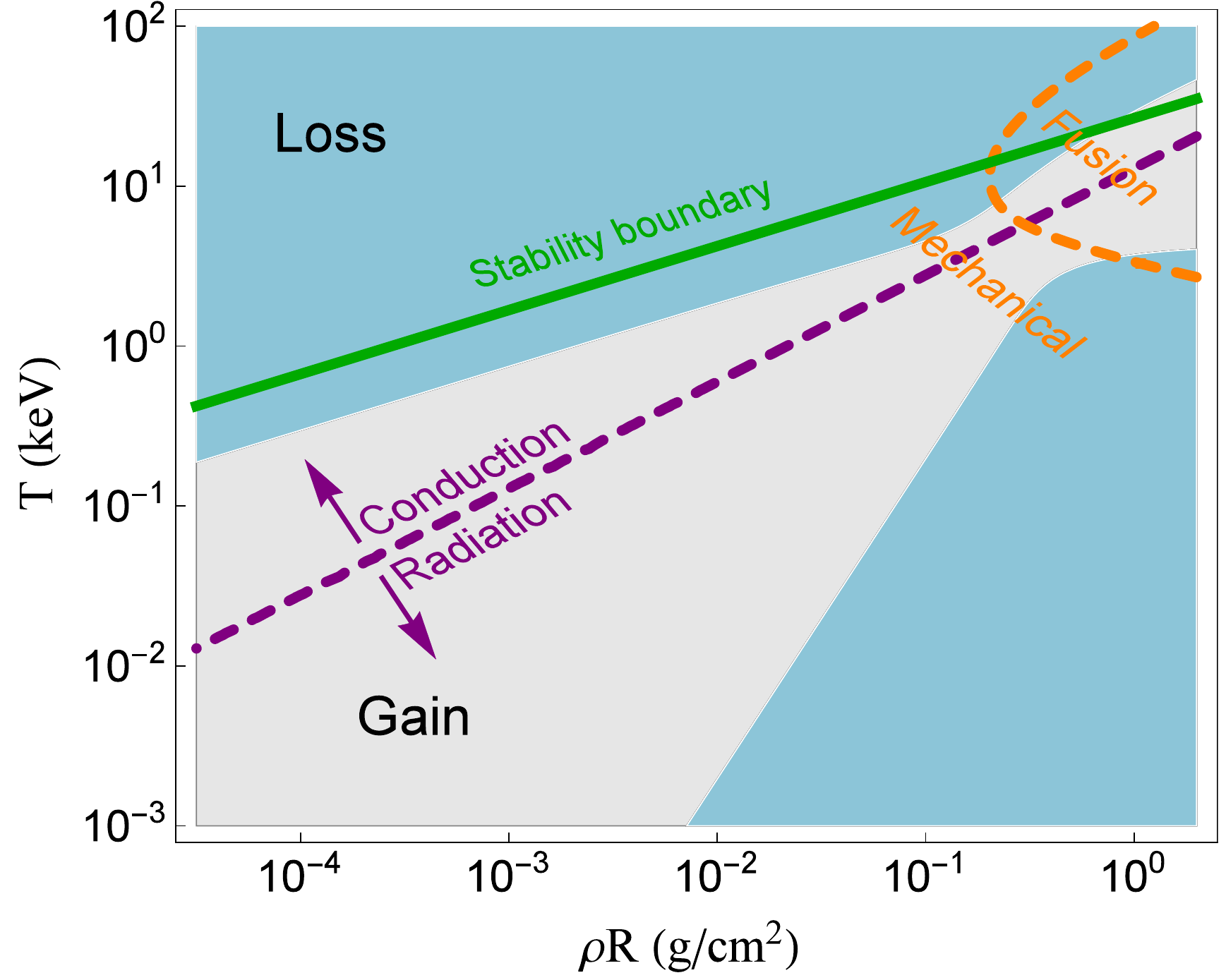}
\caption{Gain (lighter, gray) and loss (darker, blue) regions in $T,\rho R$ space using a simple hot-spot model \citep{lindl1995,atzeni2004}. These regions indicate where the hot spot gains or loses thermal energy during compression, due to the combined effects of electron thermal conduction, Bremsstrahlung radiation, mechanical ($P dV$) work, and D-T fusion heating. Also shown are dashed lines indicating the regions of $T,\rho R$ space where each heating or cooling mechanism dominates; a purple dashed line (straight) shows divides the regions where thermal conduction or radiation are the dominant loss mechanism, while an orange dashed line (curved, upper right) separates the fusion heating dominated region from the mechanical heating dominated region. The stability boundary is also shown. Of note is that the gain region is almost entirely below the stability boundary; see Sec.~\ref{sec:context}.
All plot components are drawn assuming a compression speed $U_b = 3\times10^7$ cm/s, a Coulomb logarithm $\ln \Lambda = 2$,  and for 50/50 D-T fusion, so that $A_i = 2.5$.
\label{fig:gain_loss_mechanisms_plot}}
\end{figure}

\begin{figure}
\includegraphics[width = \columnwidth]{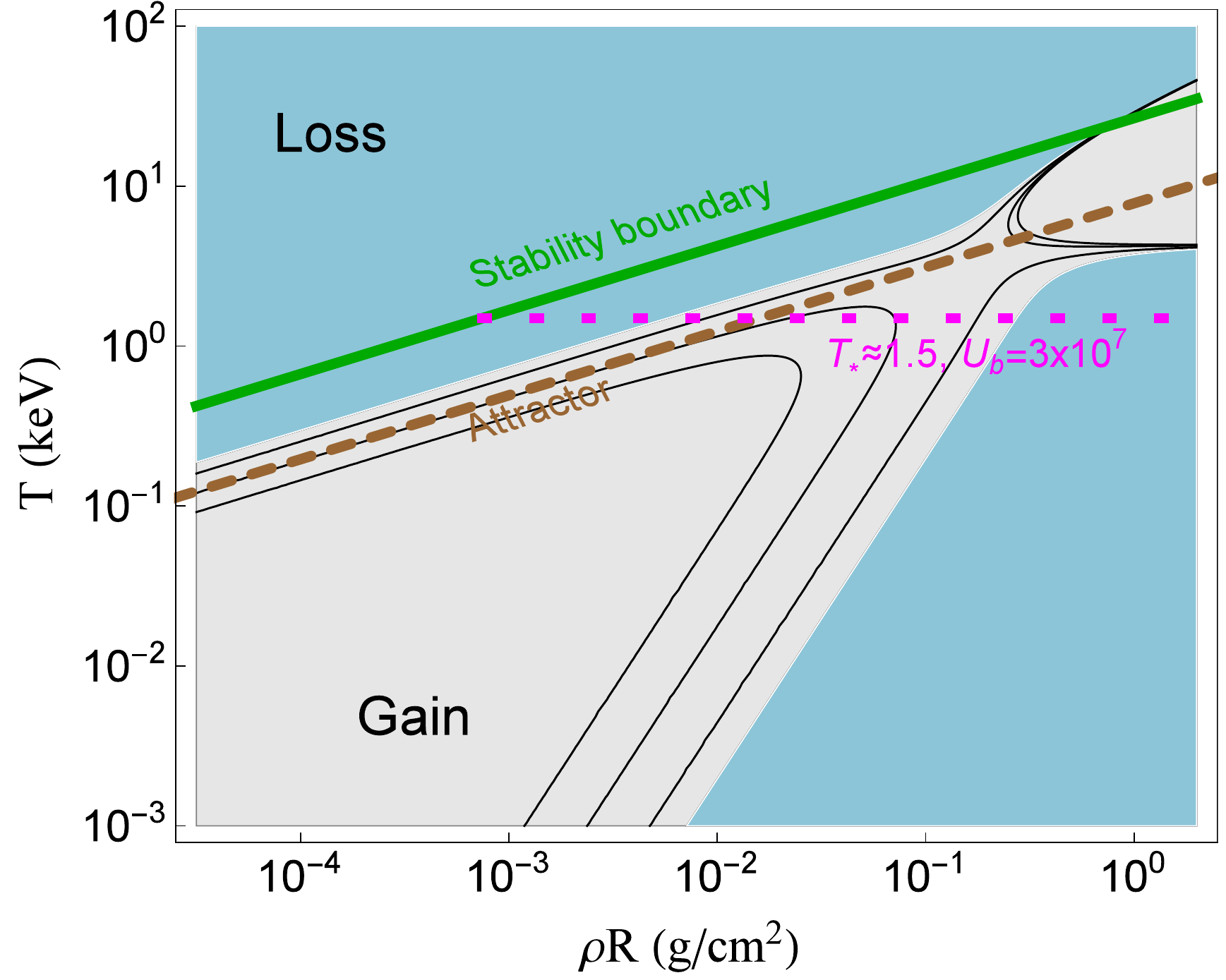}
\caption{Gain and loss regions as described in the Fig.~\ref{fig:gain_loss_mechanisms_plot} caption. Also shown is the Lindl ``attractor'' solution (brown, dashed), Eq.~(\ref{eq:lindl_attractor}), to which solutions in the simple hot-spot model are attracted. This solution is below the stability boundary, and parallels it, so that solutions following it will have TKE that tends towards $E_{\rm{sat}}$ with continuing compression. Also shown (dotted, horizontal, magenta) is the ``breakeven'' temperature, $T_{*}$, see Eq.~(\ref{eq:energy_ratio}) and the discussion thereafter, as well as Sec.~\ref{sec:context}. The thin black lines indicate show the change in the gain/loss region boundary as the compression velocity is decreased, moving from the outer-most contour ($U_b = 3\times10^7$ cm/s) to the inner-most contour ($U_b = 5\times10^6$ cm/s). Below a certain velocity, the gain region separates into two portions. The fusion-gain region can be observed to be relatively insensitive to the compression velocity; its position relative to the stability boundary then depends on the how the boundary moves with changing compression velocity (see also Fig.~\ref{fig:stability_plot}).
\label{fig:stability_with_gain_plot}}
\end{figure}

To give more context to the stability and saturation results presented in Secs.~\ref{sec:general_stability} and \ref{sec:braginskii_stability}, we consider them here paired with a simple hot-spot model. This essentially zero-dimensional hot-spot model gives the temperature of the hot-spot as a function of $\rho R$, $T_{\rm{model}}(\rho R)$. It does so by solving a temperature evolution equation that includes both heating and cooling terms. For heating terms, it includes mechanical ($P dV$) work and D-T fusion (assuming a hot-spot composed of 50/50 deuterium and tritium). For cooling terms, it includes electron thermal conduction and Bremmstrahlung radiation. We do not present the model here, it has been presented elsewhere by \citet{lindl1995} and \citet{atzeni2004}.

\begin{figure}
\includegraphics[width = \columnwidth]{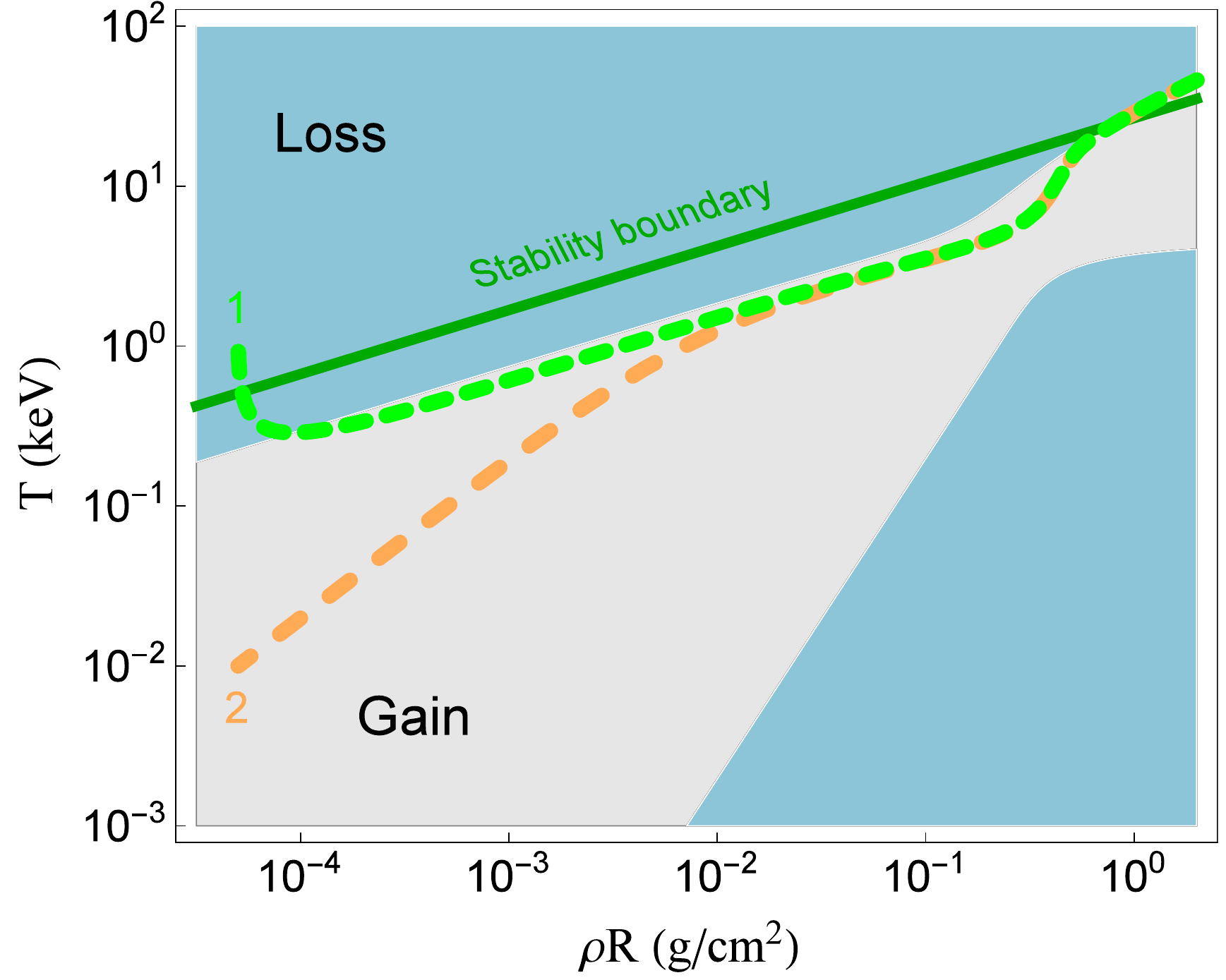}
\caption{Two example trajectories calculated for different initial conditions using the simple hot-spot model\citep{lindl1995,atzeni2004}, plotted on top of the gain and loss regions as described in the Fig.~\ref{fig:gain_loss_mechanisms_plot} caption.  One trajectory, labeled 1 (dashed, light green), starts in the guaranteed TKE decrease region, above the stability boundary. However, trajectories in this simple hot-spot model rapidly cool out of this region and head to the attractor solution. Thus while this hot-spot will briefly experience decreasing TKE (case A), it primarily exists in case B2 in Table~\ref{tbl:cases}. The second trajectory, labeled 2 (dashed, light orange), heats adiabatically until it nears the attractor solution (case B3), then parallels the attractor (case B2) into the fusion gain region. Once entering the fusion gain region, both hot-spots heat rapidly and cross the stability boundary (case A) at the margins of the gain region. All plot components use $U_b = 3\times10^7$, $\ln \Lambda = 2$, and 50/50 D-T ($A_i = 2.5$). Trajectories from more inclusive hot-spot models will have quite different behaviors, not so limited by the attractor solution, see Sec.~\ref{sec:context}.
\label{fig:example_trajectories_plot}}
\end{figure}

For a fixed compression velocity, the hot-spot model can be used to divide $T$ vs $\rho R$ space into ``gain'' and ``loss'' regions. These regions are plotted in Fig.~\ref{fig:gain_loss_mechanisms_plot}, and described in the figure caption. Also shown is an indication of the regions of $T$ vs $\rho R$ space where each loss or heating mechanism is dominant over the other loss or heating mechanism. Of note is that the portion of the gain region where mechanical heating dominates is below the stability boundary for the typical compression velocity plotted. In fact, the present hot-spot model has an ``attractor'' solution, to which trajectories that start from many initial conditions in $T,\rho R$ space will be ``attracted''. This ``attractor'' solution is valid when the heating of the capsule is determined by the balance of mechanical work and electron thermal conduction in the hot-spot model. It is given by\citep{lindl1995},
\begin{equation}
T_{\rm{attractor}} = 7.8 \left( (\rho R) \frac{U_b}{3\times10^7} \right)^{2/5} \label{eq:lindl_attractor}
\end{equation}
This result assumes $A_i = 2.5$ and $\ln \Lambda = 2$. For these values, the coefficient of the stability boundary, Eq.~(\ref{eq:stability_boundary}), is $\approx 27.0$. Then, it is easy to see that the Lindl attractor solution is below the stability boundary for any compression velocity. The attractor solution is plotted, along with the stability boundary, for $U_b = 3\times10^7$, in Fig.~\ref{fig:stability_with_gain_plot}. Also shown are contours of the gain regions for decreasing values of the compression velocity; eventually the fusion gain region becomes disconnected from the mechanical gain region, so that, within the simple hot-spot model, it becomes impossible for a trajectory to reach the fusion region.

The temperature dependence of the Lindl attractor solution, $T \propto (\rho R)^{2/5}$, is such that, for the Braginskii viscosity, it satisfies the viscosity condition for TKE saturation ($\bar{\mu} (\bar{L}) = \bar{L}^{-2}$). In other words, it satisfies condition B2 in Table~\ref{tbl:cases}, so that the TKE for solutions following the attractor will tend to grow towards $E_{\rm{sat}}$ with continuing compression. 

We can plot sample trajectories, obtained by solving for $T_{\rm{model}}(\rho R)$ with various initial conditions, $T_0,(\rho R)_0$. The expected TKE behavior of these trajectories can then be analyzed. This is done in Fig.~\ref{fig:example_trajectories_plot} and its caption. However, because of the attractor solution, this exercise does not have that many possible outcomes; trajectories can not stay in the ``stable'' region, and instead head to the attractor, on which the TKE begins to grow towards $E_{\rm{sat}}$. However, trajectories of hot-spots from experiments, or simulations with a more inclusive hot-spot model, can have quite different courses in $T$ vs $\rho R$ space, including traversing through the ``loss'' region as labeled from the simple model considered in this section. 

The gain region where fusion dominates (the lighter shaded area within the ``Fusion'' region in Fig.~\ref{fig:gain_loss_mechanisms_plot}), is less sensitive to the hot-spot model. That is, the target temperature and $\rho R$ for ignition is less dependent on the particular hot-spot dynamic model (as well as being less sensitive to the compression velocity). It is apparent from Fig.~\ref{fig:gain_loss_mechanisms_plot} that this fusion gain region is mostly in the unstable region for $U_b = 3\times10^7$. As the compression velocity decreases, it will gradually enter the TKE decrease region (see also the bottom plot in Fig.~\ref{fig:stability_plot}). At the same time, the absolute possible fraction of total energy that can be TKE in the fusion region is on order of $10\%$ at this compression velocity, with $T_{*} = 1.5$, as discussed in Sec.~\ref{sec:general_stability} and plotted in Fig.~\ref{fig:stability_with_gain_plot} (see also Fig.~\ref{fig:saturation_plot}).

\section{Discussion} \label{sec:discussion_stability}
The example trajectories shown in Fig.~\ref{fig:example_trajectories_plot}, which use the simple hot-spot model, use one compression velocity for the entirety of the compression. The present stability and saturation analysis, however, can be applied to trajectories with a compression velocity that changes; the stability boundary and $E_{\rm{sat}}$ are simply recalculated for each compression velocity. Of course, the quantitative aspects of the present analysis are not so detailed that small adjustments in $U_b$ will make a difference in the inferences made about the hot-spot TKE behavior. More sensitive analysis could be made using a TKE model\citep{davidovits2017b}. Nevertheless, the stability condition presentation here, permits gross insights into the hot-spot TKE behavior. These insights are enhanced when combined with the calculation of the saturated TKE density.

There is a reason to believe that the hot-spot TKE will have difficulty reaching $E_{\rm{sat}}$ for trajectories that satisfy case B2 (Table~\ref{tbl:cases}) but are calculated using a model that does not include TKE (such as the simple hot-spot model used for the trajectories in Fig.~\ref{fig:example_trajectories_plot}). If the TKE for such a trajectory were to grow up to be a significant fraction of the thermal energy, then the dissipation of the TKE itself would be expected to impact the trajectory. This additional heating, were it factored in, would increase the trajectory slope, effectively pushing the trajectories into case B3. The TKE can in principle be a substantial energy component as long as the temperature is less than or on order of $T_{*}$.

The present hot-spot stability and saturation results can be compared with the results of detailed three-dimensional simulations. Such simulations have been carried out for certain inertial confinement fusion experiments at the National Ignition Facility; these include an analysis of National Ignition Campaign experiment N120321 \citep{weber2015,clark2015}, which we compare to here. We focus the comparison primarily on the times before and around peak fuel velocity, since for these times the fuel-ablator interface is stable to Rayleigh-Taylor instability and the compression velocity is nearly constant (and has been for a substantial amount of compression). This compression velocity is approximated to be $\sim 3\times10^7$ cm/s. Ref.~\citenumns{weber2015} conducted both viscous and inviscid simulations of experiment N120321.

In inviscid simulations, substantial near-isotropic hydrodynamic motion is present in the hot-spot around the time of peak velocity. The saturated energy relation, Eq.~(\ref{eq:saturated_energy_density}), can be rewritten as an expression for the saturated mean fluctuating velocity, $\sqrt{\langle \mathbf{V}^2 \rangle} \approx 1.95 U_b$. Near peak velocity (22.53 ns), Ref. \citenumns{weber2015} reports burn-weighted velocity fluctuations $\sim 2.6 \times 10^7$ cm/s in the inviscid case (Fig. 6, note the plot shows individual velocity components), representing $\sim 45 \%$ of the maximum (saturated) velocity fluctuation predicted in the present work. The peak fluctuating velocities reported are $\sim 6\times 10^{7}$ cm/s (Fig. 5), approximately the mean saturated fluctuation value. It is likely that the mean fluctuating velocity is greater than $\sim 2.6 \times 10^7$ cm/s at somewhat earlier times, around 22.4 ns. This is because the viscous simulations, which show fluctuating velocities around $1.7 - 2 \times 10^7$ cm/s at peak velocity, show much higher velocities, on order of $3\times 10^7$ cm/s, or $\sim 50\%$ the saturation value, at 22.4 ns (Fig. 9), and the viscous simulations generally show fluctuation velocities that are similar to or slower than those in the inviscid simulations.

The inviscid case is very likely below the stability boundary, Eq.~(\ref{eq:stability_boundary}). While in a simulation with infinite resolution the inviscid case is necessarily below the stability boundary (as clear when taking $\mu \rightarrow 0$ in Eq.~(\ref{eq:decrease_cond})), there will generally be some numerical viscosity even when it is not explicitly included, due to the finite simulation resolution. Ref. \citenumns{weber2015} reports, for viscous simulations, a Reynolds number of $\mathrm{Re}\approx 10$ near 22.2 ns, increasing to $\approx 300$ by bang time (22.8 ns). A Reynolds number of 10 is right near the stability boundary (see Eq.~(\ref{eq:decrease_cond})); as the Reynolds number increases, the hot-spot crosses out of the TKE decrease region. In the inviscid case, the effective Reynolds number should be higher than these reported values, meaning a hot-spot below the stability boundary over the reported interval. Note that these Reynolds numbers are calculated using somewhat different length scale and velocity than those used in the TKE decrease condition, Eq.~(\ref{eq:decrease_cond}). Nevertheless, they give an approximate comparison to the condition, which is in the intended spirit.

In discussing the inviscid case above, we have effectively also covered the viscous case. Over the interval of time for which Reynolds numbers and fluctuating velocities are reported in Ref.~\citenumns{weber2015}, the viscous hot-spot is near the boundary of the TKE decrease region. Just before bang time, it has nominally moved into the ``unstable'' region; note that \emph{at} bang time, the compression velocity is no longer nearly constant. Recall that the saturated energy (or velocity) does not actually depend on the viscosity, and is therefore the same for the viscous and inviscid cases. However, the degree to which fluctuating flow in the hot-spot can reach the saturated value depends on the initial TKE, the amount of compression the hot-spot undergoes while below the stability boundary, and the viscosity growth rate during this compression. Given these factors, from the perspective of the present work, it is not surprising that the viscous hot-spot does not reach the predicted saturated values. 

The simulations in Ref.~\citenumns{weber2015} include a variety of effects that are not included in the present work, but that could influence the fluctuating velocities in the hot-spot, and therefore the validity of the comparison here. These include a jet of ablator material that enters the hot-spot after peak velocity, shocks that ring in the hot-spot, and the accretion of mass into the region defined as the hot-spot (that is, the hot-spot mass is not constant as was assumed for the analysis in the present work, and new mass, with a different fluctuating velocity condition, can be added). The degree to which these impact the ability to apply the present stability and saturation results is uncertain and would require more investigation. Much of the present comparison was carried out before the jet enters the hot-spot. To the extent the shocks (and also later, the jet) serve to seed velocity fluctuations in the hot-spot, which are then compressed, they will not necessarily negate the saturation result here. The saturation result here is most appropriate if: these sources act as a seed, this seeding is below the level of saturation, and volumetric compression remains the dominant energy injection process for velocity fluctuations. The effects on saturation of continual accretion of mass into the hot-spot depend on the velocity fluctuations in the accreted mass. If this mass is coming from a region with smaller velocity fluctuations, one would expect it to act as a damper on the hot-spot fluctuations, effectively reducing the ability of the hot-spot to reach saturation. Like the saturation result, the stability result here is also based only on the volumetric compression. As such, a hot-spot that satisfies the TKE decrease condition given here could actually experience growing TKE from shock ringing or an ablator jet. 

Note that the analysis of TKE behavior that led to the present stability and saturation results was carried out in the zero-Mach turbulence limit. This means that it may not remain valid if the hot-spot has a turbulent Mach number approaching or exceeding 1, where compressibility effects become important. Periodic boundary conditions were assumed; different boundary conditions may alter the results, to varying degrees.

These limitations mean there are a number of ways in which the present results may be refined and improved upon. This is something to be done, as our understanding of the behavior of compressing turbulence increases.

Before summarizing, we discuss the novel fast-ignition or X-ray burst scheme outlined in Refs.~\citenumns{davidovits2016a,davidovits2016b} in the context of the present stability and saturation results. These schemes propose to utilize a hot-spot (gas fill) that has its energy dominated by TKE, rather than thermal energy. Under compression, both the TKE and the thermal energy will grow. The TKE is then dissipated (through viscosity) as the $\rho R$ approaches that necessary for ignition; the dissipation of large quantities of TKE induces a rise in temperature, causing the temperature to reach that needed for ignition. By storing energy from the compression in TKE, rather than thermal energy, it is hoped energy losses, such as those to radiation, are reduced. The ratio of thermal energy to saturated TKE, eq.~\ref{eq:energy_ratio}, indicates that, for fusion, such a scheme may need compression velocities exceeding $\sim 3\times10^7$ cm/s. This is because $T_{*} \approx 1.5$ keV for this compression velocity; if the TKE reached saturation along the fast-ignition trajectory, its energy density would begin to be below that of the thermal energy once the hot-spot temperature exceeded 1.5 keV. Dissipating all the TKE at this point could at most double the thermal energy. To dissipate the TKE before fusion kicks in, the hot-spot must have a trajectory satisfying condition B3 in Table~\ref{tbl:cases} (a heating rate greater than the attractor solution). Such trajectories will usually have difficulty reaching the saturation energy, so that the ``true'' temperature at which the hot-spot thermal and TKE energy would be equal is below 1.5 keV for this compression velocity. However, we should be cautious about drawing conclusions about the proposed fast-ignition scheme from the present work, because the proposed scheme operates far from the zero-mach limit of the current treatment.

\section{Summary} \label{sec:summary_stability}
We present a what we call a ``stability'' condition for hot-spot turbulent energy; if satisfied, this condition guarantees the turbulent (non-radial hydrodynamic) energy in the hot-spot will decrease while the hot-spot undergoes compression. When it is not satisfied, the TKE behavior depends on the precise hot-spot trajectory and conditions, but can in many cases be bounded by a saturated hot-spot turbulent energy (density) $E_{\rm{sat}}$. We calculate this saturated value. The stability boundary is shown visually for the case where the hot-spot has Braginskii viscosity, and we describe how to determine, visually, cases where the hot-spot TKE can be bounded by $E_{\rm{sat}}$. By comparing the saturated turbulent energy to the hot-spot thermal energy, we determine the maximum fraction of hot-spot energy that can be TKE for any given compression velocity.

We show that trajectories in a simple hot-spot model will quickly enter the ``unstable'' TKE region, and that most of the ``gain'' region for hot-spot thermal energy in this model is below the stability boundary. These trajectories largely follow an ``attractor'' solution, which has TKE that will grow towards $E_{\rm{sat}}$ with continuing compression.

We hope this theoretical perspective captures the gross behavior of hot-spot turbulence.

\begin{acknowledgments}
This work was supported by NNSA 67350-9960 (Prime $\#$ DOE DE-NA0001836), by NSF Contract No. PHY-1506122, and by the U.S. Department of Energy Fusion Energy Sciences Postdoctoral Research Program administered by the Oak Ridge Institute for Science and Education (ORISE) for the DOE. ORISE is managed by Oak Ridge Associated Universities (ORAU) under DOE contract number DE-SC0014664. All opinions expressed in this paper are the authors' and do not necessarily reflect the policies and views of DOE, ORAU, or ORISE.
\end{acknowledgments}

\end{document}